\documentclass[12pt,titlepage]{article}

\usepackage{graphicx}
\usepackage{color}
\usepackage{xspace}

\renewcommand{\theequation}{\thesection.\arabic{equation}}

\font\medio=cmr10 scaled \magstep2
\outer\def\beginsection#1\par{\medbreak\bigskip
      \message{#1}\leftline{\bf#1}\nobreak\medskip
\vskip-\parskip
      \noindent}

\def\laq{\raise 0.4ex\hbox{$<$}\kern -0.8em\lower 0.62
ex\hbox{$\sim$}}
\def\gaq{\raise 0.4ex\hbox{$>$}\kern -0.7em\lower 0.62
ex\hbox{$\sim$}}

\def\beq{\begin{equation}}
\def\eeq{\end{equation}}
\def\bea{\begin{eqnarray}}
\def\eea{\end{eqnarray}}

\def \pa {\partial}

\def \fb {\overline \phi}
\def \fbp {\dot{\fb}}
\def \bp {\dot{\beta}}

\def \ti {\tilde}
\def \la {\lambda}

\def \b {\beta}
\def \a {\alpha}
\def \ap {\alpha^{\prime}}

\def \ga {\gamma}
\def \sg {\sigma}
\def \da {\delta}

\begin{document}
\bibliographystyle {unsrt}

\titlepage
\begin{flushright}
CERN-TH/96-267 \\
IFUP-TH/62-96\\
hep-th/9611039\\
\end{flushright}

\begin{center}
{\bf TOWARDS A NON-SINGULAR  PRE-BIG BANG COSMOLOGY}\\

\vspace{9mm}

M. Gasperini\footnote{Permanent address:
{Dip. Fisica Teorica, Un. di  Torino, Via P. Giuria 1, 10125 Turin,
Italy.}} \\
{\sl Theory Division, CERN, CH-1211 Geneva 23, Switzerland} \\
M. Maggiore\\
{\sl Istituto Nazionale di Fisica Nucleare and Dipartimento di Fisica
dell'Universit\`a, \\
Piazza Torricelli 2, I-56100 Pisa, Italy} \\
and\\
G. Veneziano \\
{\sl Theory Division, CERN, CH-1211 Geneva 23, Switzerland} \\
\end{center}
\vspace{9mm}
\centerline{\medio  Abstract}

\noindent
 We discuss  general features of
the  $\beta$-function equations for  spatially flat,
$(d+1)$-dimensional  cosmological backgrounds at lowest order in the
string-loop expansion,
 but to all orders in $\a'$.  In the special case of
constant curvature and a linear dilaton these equations
 reduce to $(d+1)$ algebraic
equations in $(d+1)$ unknowns, whose solutions
 can  act as late-time regularizing attractors for the
singular lowest-order
pre-big bang solutions. We illustrate the phenomenon in a
first order  example, thus
providing an explicit realization of the previously conjectured
transition from the dilaton to the string phase in the weak
coupling regime of string cosmology. The complementary role of
$\ap$ corrections and string loops for  completing the transition
to the standard cosmological scenario is also briefly discussed.

\vspace{8mm}

\centerline{Published in {\bf Nucl. Phys. B 494, 315 (1997)}}


\vfill
\begin{flushleft}
CERN-TH/96-267 \\
September 1996
\end{flushleft}

\newpage

\renewcommand{\theequation}{1.\arabic{equation}}
\setcounter{equation}{0}
\section {Introduction}

Standard cosmology \cite{1} assumes that the
primordial Universe was in a hot,
dense, and highly curved state, very close indeed to the so-called
big-bang singularity. By contrast, the duality symmetries of string
theory \cite{2,3,4} motivate a class of ``pre-big bang" cosmological
models \cite{3,5} in which the Universe starts very near the cold, 
 empty and flat perturbative vacuum.
This scenario, although theoretically appealing,
can only make sense phenomenologically if such unconventional
initial conditions evolve
naturally into those of the standard scenario at some later time,
smoothing out the big-bang singularity. Does this happen?

The early evolution of a Universe that starts at very low curvature
and coupling is well described by the low-energy, tree-level string
effective action \cite{6,6a}. The corresponding field equations imply
that the string perturbative vacuum, with vanishing coupling constant
$g_s= e^{\phi/2}=0$,  is actually  unstable towards
 small homogeneous fluctuations of the metric and the dilaton $\phi$, 
which can easily ignite an accelerated growth of the curvature and of
the coupling \cite{3,5}. However, in the absence of higher-order
corrections to the effective action, such a growth is unbounded:
as conjectured in \cite{7} (and later proved to a large extent in
\cite{8}), a
singularity in the curvature and/or the coupling is reached in a finite
amount of cosmic time, for any realistic choice of the (local) dilaton 
potential.
Is such a singularity unavoidable?

Higher-order corrections to the effective action of string theory are
controlled by two independent expansion parameters.
One is the field-dependent (and thus in principle
space-time-dependent) coupling $g_s$,
which controls the importance of string-loop corrections. 
The other parameter, $\ap$,  controls the
importance of finite-string-size corrections, which are small if fields
vary little over a string-length distance $\lambda_s = \sqrt{\a'}$. 
Only when this second expansion parameter is small, can the
higher-derivative corrections to the action  be
neglected and does string theory go over to an
effective quantum field theory.

As discussed in previous examples \cite{10}-\cite{13}, quantum
corrections arising in the strong coupling regime can regularize the
curvature singularity of the tree-level pre-big bang models, already at
the one-loop order. With the exception of the very particular case of
two space-time dimensions, the inclusion of loops is necessarily
associated with the appearance of higher-derivative terms in the
effective action, and thus requires also,
for consistency, the inclusion of
higher orders in $\ap$. Instead, if the initial value of
the string coupling is small enough, it is quite possible that the 
Universe reaches the high-curvature regime in which
higher-derivative ($\ap$) corrections become important, while the
coupling is still small enough to neglect loop corrections.

According to the above considerations, we shall discuss
hereafter the  regularization of the
curvature singularity  just as a   result of
``stringy" $\ap$ corrections, but at lowest order in $g_s$. 
We will consider, in particular, the
possibility that a cosmological background evolving from the
perturbative vacuum
be attracted into a state of constant curvature and linearly running
dilaton, i.e. of constant $H$ and $\dot\phi$  ($H=\dot a/a$ is the Hubble
parameter in the so-called string frame, i.e. in the frame most directly
related to the $\sg$-model parametrization of the action, and
the dot stands for
differentiation with respect to cosmic time in that frame).

A high-curvature string phase with frozen  string-size values of  $H$
and $\dot{\phi}$ was previously conjectured as a crucial ingredient for
a successful pre-big bang scenario, and
has important phenomenological consequences \cite{16}.
Such a state exists in general as a cosmological solution
of higher-curvature effective actions, but is in general
disconnected from the perturbative vacuum (the trivial solution with
$H=\dot\phi=0$) by a singularity, or by an unphysical region in which
$ H$ becomes imaginary. String theory, on the contrary, provides
examples in which the two states are smoothly joined by the
evolution of the background already to first order in $\ap$, as we
will show in this paper. In this
sense string theory automatically implements, without a scalar
potential and in any number of dimensions, the ``limiting curvature"
hypothesis previously introduced ad-hoc, with various mechanisms
\cite{14,15}, to regularize the curvature of cosmological backgrounds.

A final state with linearly evolving
dilaton can also be obtained in the context of
one-loop-regularized models. The main difference between the
 loop case and the one at hand is
that our  final shifted
dilaton $\fb$ satisfies, in $d$ isotropic spatial dimensions, 
$\fbp \equiv \dot
\phi -dH<0$. This is a necessary condition for the background to be
subsequently attracted by an appropriate potential in a state with
expanding metric ($H>0$) and frozen dilaton ($\dot\phi=0$), the starting
point of standard cosmology.
In fact, since the dilaton keeps growing after the transition to the
string phase, the effects of loops, of a non-perturbative dilaton
potential, and of the back-reaction from particle production must
eventually become important, and are
expected to play an essential role in the second transition from the
string phase to the usual hot big bang scenario.
Leaving this second transition to further investigation, the main
purpose of this paper is to give arguments in favour of
the occurrence of a smooth evolution, in the weak 
coupling regime, from
the dilaton phase to the constant curvature string phase dominated by
the $\ap$ corrections.

The paper is organized as follows.
In Section 2 we discuss the general structure of tree-level, 
$\sg$-model $\beta$-functions for generic Bianchi-type I backgrounds
(including a time-dependent dilaton). In Section 3 we specialize the
equations to the case of constant curvature
and linear dilaton, and show that
in this case we obtain a system of $(d+1)$ algebraic equations in
$(d+1)$ unknowns ($d$ Hubble constants and $\dot{\phi}$). The
equations  look like those for the fixed points of a renormalization
group (RG) flow in physical cosmic time (as opposed 
to the RG ``time"). In Section 4 we consider an example to first order in
$\ap$ (i.e. four derivatives), determine the fixed points, and show,
by numerical integration, that any isotropic pre-big bang background
necessarily evolves smoothly  towards the regular fixed
points, thus avoiding the singularity. The fixed points have in general an
attraction basin of finite area, in the space of initial conditions, also for
anisotropic backgrounds.  Section 5 contains our conclusions.

\renewcommand{\theequation}{2.\arabic{equation}}
\setcounter{equation}{0}
\section { Bianchi I tree-level $\beta$-functions}

It is well known \cite{6,6a} that the field equations for the background
fields appearing in the $\sigma$-model action of string theory
correspond to the vanishing of a set of $\beta$-function(al)s. We shall
call the latter  ``$\sigma$-model $\beta$-functions" in order to avoid a
possible  confusion with another set of
effective  $\beta$-functions to be introduced later. 
It is also known \cite{6, 6a} that
the $\sigma$-model $\beta$-functions vanish on a set of
field equations that can be
derived by varying a space-time effective action $\Gamma$, which is
also the generating  functional of the string $S$-matrix.

We shall consider in this paper a $\sigma$-model  background
in ($d+1$) dimensions, consisting  of just a time-dependent dilaton
$\phi(t)$ and  an anisotropic Bianchi-type I metric, $g_{\mu\nu}$,
which can be conveniently parametrized as:
\beq
g_{\mu\nu} = {\rm diag} \left(N^2(t), -
a_i^2(t)\da_{ij}\right) ; ~~~~~~~~i,j = 1,... ,d .
\label{21}
\eeq
Here $N(t)$
is the lapse function and $a_i(t)$ are the scale factors along
the $d$ different spatial
directions. We will not consider, for the sake of simplicity, an additional
antisymmetric-tensor background $B_{\mu\nu}$, which is needed in
order to discuss the $O(d,d)$ symmetries \cite{19,20} associated with
the presence of $d$ Abelian isometries, but our considerations can 
easily be extended to such a case.

Using the general covariance of the string effective action  and
the fact that, at tree-level in the string-loop expansion, the dependence
on a constant dilaton is fixed in the string frame,
we can immediately write the exact,
tree-level effective action for this background, in terms of the fields
\beq
\fb =\phi - \sum_i \beta_i, ~~~~~~~~~~
\beta_i = \ln~ a_i ,
\label{22}
\eeq
 in the form:
\beq
\Gamma =  \int dt N e^{-\fb}
{\rm L}\left(\bp_i^{(n)}, \fbp_i^{(n)}\right).
\label{23}
\eeq
The effective Lagrangian $\rm{L}$ is a general function of the (properly
covariantized) time derivatives of
the fields $\b_i$ and $\fb$, at all orders in $n\geq 1$:
\beq
\bp_i^{(n)}=\prod_{k=1}^n\left({1\over N}{d\over dt}\right)^k 
\b (t),~~~~~~~~
\fbp^{(n)}=\prod_{k=1}^n\left({1\over N}{d\over dt}\right)^k 
\fb (t) .
\label{24}
\eeq
In this paper we shall limit our attention to the case of critical
(super)strings in which no cosmological constant appears in ${\rm L}$. 
Thus, even when we discuss the case $d\not= d_{{\rm crit}}$ we shall
assume that other  ``passive" sectors are present to cancel the 
central charge deficit $(d-d_{{\rm crit}})/3 \ap$.

There are just three kinds of $\sigma$-model $\beta$-function
equations that follow from
(\ref{23}).  They correspond, respectively, to the field equation for the
lapse,  for each scale factor, and for the  dilaton $\fb$.
However, these ($d+1$)
equations are not independent. The relation among
them follows from the general fact that, in a theory with coordinate
reparametrization invariance, the set of ``non-dynamical" field
equations representing constraints on the initial data are covariantly
conserved \cite{1}, as a consequence of the Bianchi identities. In our
particular case, the residual
time reparametrization invariance implies the relation:
 \beq
\dot{\beta_i} {\delta \Gamma  \over \delta {\beta_i} } \; +
\fbp {\delta \Gamma  \over \delta {\fb} }
 = N {d \over dt}{\delta \Gamma \over \delta N} .
\label{25}
\eeq
On the equations of motion each one of the three terms in eq.
(\ref{25}) vanishes separately, so that only $(d+1)$ equations are
independent, to all orders, in agreement with a general property of the
$\sg$-model $\b$-functions \cite{21}.

The variation of the action (\ref{23}), in the cosmic time gauge
$N=1$, gives rise to the following system of field
equations:
 \beq
 {\partial {\rm L} \over \partial \dot{\beta_i} } -
e^{\fb} {d \over dt}\left (e^{-\fb} {\partial {\rm L}
\over \partial \ddot{\beta_i} }\right) +
\dots = e^{\bar{\phi}} Q_i  ,
\label{26a} 
\eeq
\beq
  -\rm{L} - e^{\fb} {d \over dt} \left(e^{-\fb}
{\partial \rm{L} \over \partial \fbp}\right)+
e^{\fb} {d^2 \over dt^2} \left(e^{-\fb}
{\partial \rm{L} \over \partial \ddot{\fb}}\right) + \dots = 0\; ,
\label{26b}
\eeq
\beq
 {\rm L} -  \fbp {\partial {\rm L} \over \partial
\fbp}
- \dot{\beta_i}{\partial {\rm L} \over \partial \dot{\beta_i}}
 -2 \ddot{\beta_i} {\partial {\rm L} \over \partial \ddot{\beta_i}}
 -2 \ddot{\fb} {\partial {\rm L} \over \partial \ddot{\fb}} +
e^{\bar{\phi}} {d \over dt} \left(e^{-\bar{\phi}} \dot{\beta_i}
{\partial {\rm L} \over \partial \ddot{\beta_i}}+
e^{-\bar{\phi}} \fbp
{\partial {\rm L} \over \partial \ddot{\fb}}
\right) + \dots = 0 .
\label{26c}
\eeq
Equation (\ref{26a}), obtained by varying $\beta_i$, 
defines a set of $d$  charges $Q_i$ whose conservation
follows from the fact that $\Gamma$ depends
only upon derivatives of the $\beta_i$. The second equation
is the (shifted) dilaton
equation, while the last equation is the so-called Hamiltonian
constraint, following from the variation of the lapse. Relation
(\ref{25}) can be directly checked at the level of  eqs.
(\ref{26a})--(\ref{26c}).

 Equation (\ref{25}) can be exploited in several ways when solving
the system (\ref{26a})--(\ref{26c}). The first way, which  we  
adopt here,  is to solve
the first two equations and to impose the constraint only on the initial
data. After solving the equations numerically, we can double check that
the constraint remains valid at all times. Alternatively, we can  just use
the constraint and the
 conservation equations and ignore the dilaton equation.
Equation (\ref{25}) then
guarantees that the dilaton equation is automatically satisfied provided
$\fbp$ is not identically zero.

\renewcommand{\theequation}{3.\arabic{equation}}
\setcounter{equation}{0}
\section { The constant curvature case}

Let us now consider a very special class of backgrounds, those with
constant curvature and a linear dilaton, $\bp_i$ and $\fbp$ constant,
where the fields are referred to the so-called  string frame defined
by the action (\ref{23}).  In our coordinate
system they correspond to the ansatz:
\bea ds^2  = dt^2 - \sum_i e^{2 H_i t}
dx^i dx^i , ~~~~~~~ \phi(t) = c t + \phi_0
\label{31}
\eea
parametrized by the $(d+1)$ constants  $c$ and $H_i$.

The only known (all-order) solutions of this type  have a trivial
Minkowski metric and  a constant or linear dilaton \cite{22} depending
on whether $d = d_{\rm{crit}} $ or $d > d_{{\rm crit}}$.
The existence of other
high-curvature solutions of this type  is all but excluded \cite{23},
however, provided the dilaton is not a constant, $c\not=0$. Note,
incidentally, that it is quite crucial that one looks at constant curvature
solutions in the string frame \cite{23}. 
Because of the dilaton's time-dependence
this is {\underline {not}}
equivalent to a constant curvature in the Einstein frame,
for which no-go theorems probably apply \cite{24}. Let us thus discuss
the necessary and sufficient conditions for the existence of solutions of
this type, which, in accordance with  previously used terminology
\cite{16}, we will term string-phase solutions.

 It is clear, by inspection, that, for this class of
backgrounds, eqs. (\ref{26a})--(\ref{26c}) reduce to $(d+2)$
algebraic equations in the $(d+1)$
unknowns $H_i$ and $c$. However, as already discussed, 
only $(d+1)$ equations
are really independent, giving rise to
the hope that isolated solutions other than the already
mentioned trivial ones might exist.
Actually, from   eq. (\ref{26a}),
we easily see that the l.h.s.
is constant for a string-phase solution. Thus, in order for the r.h.s.  
to be
constant as well, two possibilities exist: either $\fb$ itself is
constant with  $Q_i$ arbitrary, or $Q_i = 0$ for all $i$.
 Let us discuss these two possibilities in turn: \begin{itemize}
\item[1)] $\fbp \equiv 0$.

This case looks more attractive at first sight. However,
with $\fbp \equiv 0$, the remaining two equations
are independent (see
discussion in the previous Section) and give two constraints
among the $d$ conserved charges
$Q_i$.
Since these charges  are already related by the initial (pre-big bang)
conditions, fine-tuned initial conditions would be needed in order to
flow into a string phase of this kind. For this reason we will
concentrate here on the second alternative.
\item[2)] $\fbp\not= 0, \;  Q_i =0$.

This case looks at first even worse because, instead of having two
constraints among the $d$ conserved charges, we now have  
set all of them to zero while they are certainly non-vanishing on the
pre-big bang solution. However, if  $\fbp < 0$, both sides of
eq.(\ref{26a}) will go exponentially to zero at late times, irrespectively
of the $Q_i$'s. Therefore, as we shall see
explicitly in the following section, string-phase solutions
with $\fbp < 0$ can play the
role of late-time attractors for solutions coming
from pre-big bang initial conditions.

\end{itemize}
Although $\fbp < 0$ is a necessary condition for the above phenomenon
to  occur, it turns out not to be always sufficient. This can be most
easily understood by using a RG description. Our
differential equations (\ref{26a})--(\ref{26c}) define a set 
of RG equations in
physical (cosmic) time where $\fbp$, $\dot{\beta_i}$ play the
role of running couplings. Time-derivatives of the couplings define some
new kind of $\beta$-functions whose zeros correspond to the
constant-curvature, linear-dilaton fixed points. Trivial Minkowski space
(with $\fbp=0$) corresponds to a trivial quadratic zero of the
$\beta$-functions and, consequently, is a late-time (early-time)
attractor  for post-big bang (pre-big bang) initial conditions.

By contrast, a non-trivial simple zero with
$\dot{\bar{\phi}} < 0$ is a late-time attractor
from any initial condition sufficiently close to it.
In order that pre-big bang initial
conditions flow to this attractor, it is necessary
that no other zeros or singularities
separate the trivial fixed point from the non-trivial one.
If this is the case, the
long-conjectured transition from the dilaton
to the string phase does indeed take place.
In the next section we shall see explicit examples of such
an interesting phenomenon.

\renewcommand{\theequation}{4.\arabic{equation}}
\setcounter{equation}{0}
\section {A first-order example}

To the first order in $\ap$, and in the string frame, the simplest
effective action that reproduces the massless bosonic
sector of the tree-level string
$S$-matrix can be written in the form \cite{6a}:
\beq
S=-{1\over 2\la_s^{d-1}}\int d^{d+1}x \sqrt{|g|}e^{-\phi} \left[ R+
(\nabla \phi)^2-{k\ap \over 4} R_{\mu\nu\a\b}^2\right]\; ,
\label{41}
\eeq
where $k=1,1/2$ for the bosonic and heterotic string, respectively
(remember that we have assumed  the torsion background to be 
trivial). The
corresponding field equations are equivalent, in the sense discussed in
\cite{6a}, to the conditions of $\sg$-model conformal invariance
represented by the vanishing of the $\b$-functions.

In order to discuss cosmological solutions, it is convenient to perform a
field redefinition (preserving, however, the $\sg$-model 
parametrization
of the action) that eliminates terms with higher than second
derivatives from the effective equations. This can be easily done, as
is well known, by replacing the square of the Riemann tensor with the
Gauss--Bonnet invariant \cite{25} $R^2_{GB} 
\equiv R_{\mu\nu\a\b}^2-4  R_{\mu\nu}^2+
R^2$,  at the price of introducing dilaton-dependent $\ap$
corrections. The field redefinition
\beq
\ti g_{\mu\nu}=g_{\mu\nu}+4k \ap \left[
R_{\mu\nu}-\pa_\mu\phi\pa_\nu \phi+ g_{\mu\nu}(\nabla
\phi)^2\right], ~~~
\ti \phi =\phi +k\ap \left[R+(2d-3)(\nabla\phi)^2\right],
\label{42}
 \eeq
truncated to first order in $\ap$,
leads in particular to the following simple form of the action (dropping
the tilde over the redefined fields):
\beq
S=-{1\over 2\la_s^{d-1}}\int d^{d+1}x \sqrt{|g|}e^{-\phi} \left[ R+
(\nabla \phi)^2-{k\ap \over 4} \left(R^2_{GB} - 
(\nabla \phi)^4\right)\right] ,
\label{43}
\eeq
which we will use throughout this section.

We have explicitly checked
that the results of this section are  qualitatively
reproduced also by using different parametrizations of the string
effective action, for instance the one obtained by transforming 
into the string frame a pure Gauss--Bonnet action in 
the Einstein frame. Such results are not invariant, however, under field
redefinitions that are truncated by keeping  first order terms 
in $\ap$ in the effective action. With an additional truncated
redefinition we can in fact modify eq.(\ref{43}), without re-indroducing
second derivatives, and  obtain, in particular, the special action that
was shown to be off-shell equivalent to the conditions of $\sg$-model
conformal invariance \cite{26a} through  Zamolodchikov-like equations.
 The same action has also been recently proposed as the correct one 
 to achieve a higher-order
extension of the T-duality symmetry \cite{26b}. 
Such an action does have fixed
points; however, these are not smoothly
connected to the perturbative vacuum. This
field-redefiniton dependence of the background properties 
is unavoidable
as long as the $\ap$ expansion is truncated at a given finite order.

We specialize the Bianchi I background, for simplicity, to the case in
which the spatial sections are the product of two isotropic, conformally
flat (``external" and ``internal") manifolds, respectively $d$- and
$n$-dimensional, described by the metric
\beq
g_{00}=N^2(t), ~~~~g_{ij}=-\da_{ij} e^{\b(t)}, ~~~~
g_{ab}=-\da_{ab} e^{\ga(t)}, ~~~~i,j=1,...,d ;~~~~ a,b=d+1,...,d+n
\label{44}
\eeq
(note that the total number of spatial dimensions, previously 
denoted by $d$, is now redefined to be  $d+n$).
After integration by parts, the action (\ref{43}) can be written as
$$
 S \propto \int dt e^{d\b+n\ga-\phi} \left[{1\over
N}\left(-\dot\phi^2 - d(d-1)\bp^2-n(n-1)\dot \ga^2-2dn\bp \dot
\ga +2d\bp\dot\phi+2n\dot \ga\dot\phi\right)+\right.
$$
\beq
\left. {k\ap\over 4 N^3}\left(c_1\bp^4+c_2\dot \ga^4+
c_3\dot\phi\dot \b^3+c_4\dot\phi\dot\ga^3+c_5\dot\phi\dot \b
\dot\ga^2+c_6\dot\phi\dot \b^2\dot\ga+
 c_7\dot \b^2\dot\ga^2+
c_8\dot \b\dot\ga^3+c_9\dot \b^3\dot\ga-\dot\phi^4\right)\right]\; ,
\label{45}
\eeq
where
\bea
c_1&=&-{d\over 3}(d-1)(d-2)(d-3), ~~c_2=-{n\over 3}(n-1)(n-2)(n-3),
\nonumber \\
c_3&=&{4\over 3}d(d-1)(d-2), ~~
c_4={4\over 3}n(n-1)(n-2),
~~c_5=4dn(n-1), ~~
c_6=4dn(d-1),\nonumber \\
c_7&=&-2dn(d-1)(n-1), ~c_8=-{4\over 3}dn(n-1)(n-2), ~
c_9=-{4\over 3}dn(d-1)(d-2) .
\label{46}
\eea
Here we are working, for convenience, with the original dilaton
$\phi$, but the action can be easily converted to the general form of
section 2 using the definition  $\phi \equiv \fb+d\b+n\ga$.

Let us first look for constant curvature solutions in the isotropic case
$n=0$. By varying the action with respect to $\phi$ and $N$, and setting
$\dot\phi=x={\rm const}$, $\bp=y={\rm const}$, in the gauge $N=1$, we
get the two independent algebraic equations
\bea
x^2&+&d(d-1)y^2-2dxy-{k\ap\over 4} \left(c_1
y^4+c_3xy^3-x^4\right)-\nonumber \\
&-&(dy-x)\left[-2x+2dy+{k\ap\over 4}\left(c_3y^3-4x^3\right)
\right]=0,
\nonumber \\
x^2&+&d(d-1)y^2-2dxy-{3\over 4}k\ap \left(c_1 y^4+c_3xy^3-x^4
\right)
=0.
\label{47}
\eea
We have explicitly checked that they have
real solutions for any $d$ from $1$ to $9$.
For $d=3,6,9$, in particular, the
coordinates of the fixed point in the plane $(\dot\phi, \bp)$ are given by
(in units $k\ap=1$)
\bea
d&=&3 ~~~~~~~~~~x=\pm1.40..., ~~~~~~y=\pm0.616...,\nonumber\\
d&=&6 ~~~~~~~~~~x=\pm1.37..., ~~~~~~y=\pm0.253...,\nonumber\\
d&=&9~~~~~~~~~~x=\pm1.38..., ~~~~~~y=\pm0.163..., ~~.
\label{48}
\eea
Exactly the same results follow from the system of equations obtained
by varying $\{N, \b\}$ and $\{\phi, \b\}$. In the last case one 
gets an additional fixed point, which corresponds, however, to
$\fbp=x-dy=0$. In that case the third equation is no longer a
consequence of the other two: indeed,  by imposing the constraint $\da
S/\da N=0$, one finds  that the additional solution has to be discarded,
in agreement with the general discussion of the previous sections.
Although we have not made an exhaustive search, it seems that
non-isotropic fixed points are excluded in the absence of an exact
duality symmetry.

By integrating numerically the field equations for $\b$ and $\phi$, and
imposing the constraint on the initial data, we have verified that, for
any given initial condition corresponding to a state of pre-big bang
evolution from the vacuum (i.e. $0<\bp<x$, $\fbp=\dot\phi-d\bp >0$), 
the
solution is necessarily attracted to the expanding fixed points
(\ref{48}). This is illustrated in Fig. 1, for various numbers of dimensions.

\begin{figure}[t]
\centerline{\includegraphics[scale=0.8]{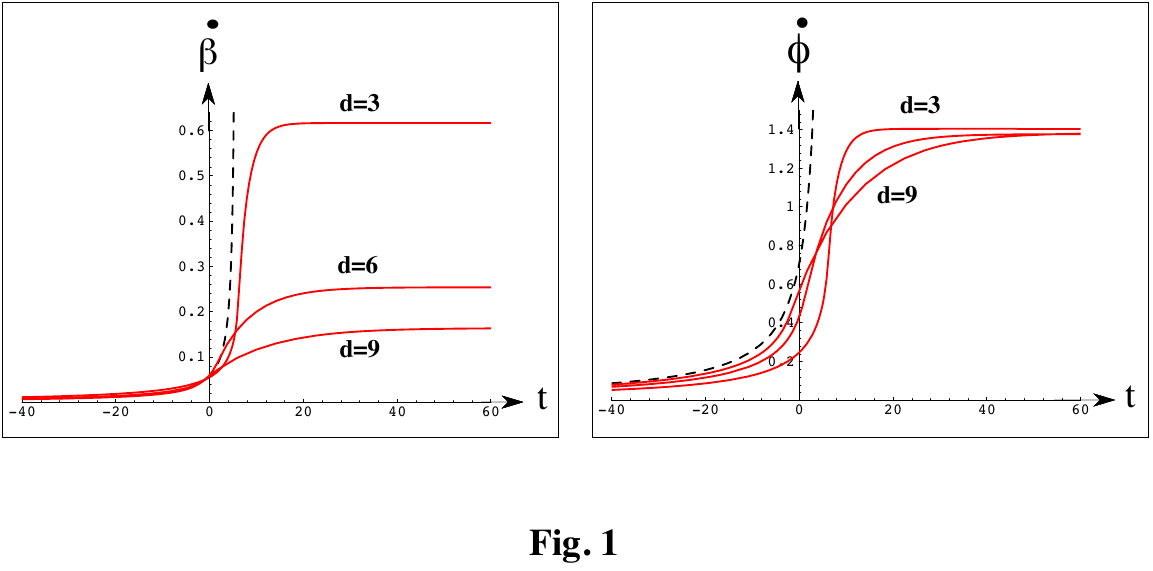}}
\caption{Curvature regularization of a pre-big bang background as a
consequence of the first-order $\ap$ corrections (in units $k\ap=1$). The
dashed curve shows the singular behaviour of the zeroth-order solution
in $d=9$. The solid curves approach asymptotically the constant values 
of eq. (\ref{48}).}
\label{f1}
\end{figure}


In spite of the fact that the fixed points generically appear when
higher curvature terms are added to the action,  they
cannot always be reached from the perturbative vacuum, because 
of an intermediate
singularity or of an unphysical, classically impenetrable region. This is
what happens, for instance, if one considers the previously 
discussed action with
the Gauss--Bonnet term alone, which does not correspond,
in the string frame, to the correct string effective action. The
Gauss--Bonnet invariant, by itself, may parametrize the 
first-order $\ap$
corrections only in the Einstein frame, and only in $d=3$ \cite{6a}. 

For the string effective action (\ref{43}), on the contrary, the fixed
points are continuously joined to the perturbative vacuum
($\bp=0=\dot\phi$) by the smooth flow of the background in cosmic
time, as illustrated by the ``$\b$-functions" $\ddot\b (\bp)$,
$\ddot\phi (\dot\phi)$ plotted in Fig. 2 
($\dot\phi$ and $\dot\b$ are obviously not 
independent, being related by the constraint equation). 
They show the running of the
curvatures $\bp$, $\dot\phi$ for the numerical solutions of Fig. 1. The
simple action ({\ref{43}) thus implements, already to first order
in $\ap$, a smooth transition from the dilaton phase to the string phase
of the pre-big bang scenario, in agreement with previous assumptions
\cite{16}.

\begin{figure}
\centerline{\includegraphics[scale=0.8]{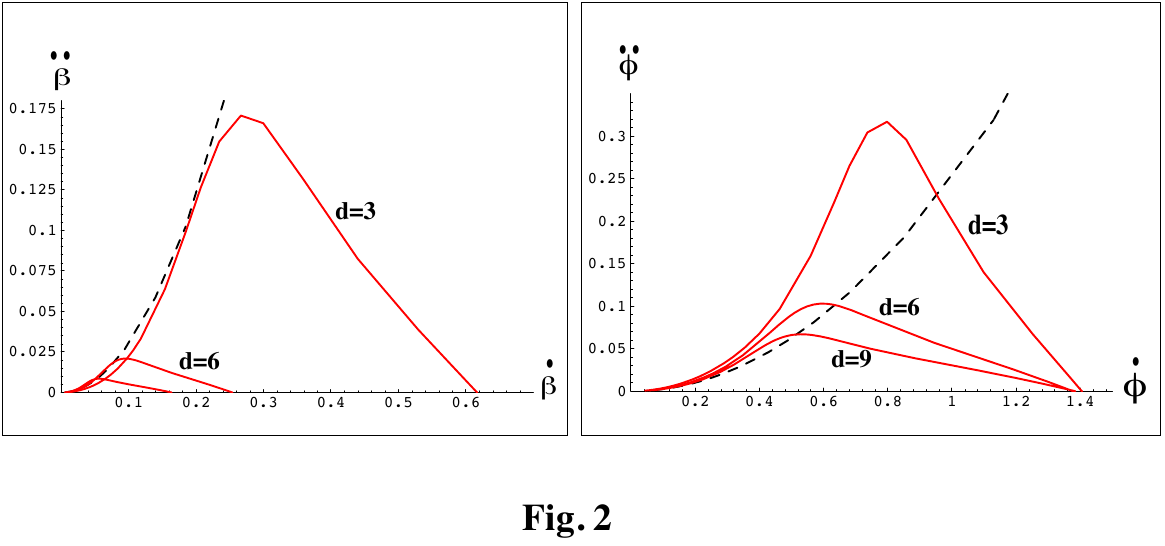}}
\caption{Smooth evolution of the background from the perturbative
vacuum, $\bp=0=\dot\phi$, to the non-trivial fixed points with $\bp$
and $\dot\phi$ constant reported in eq. (\ref{48}). 
The dashed curve corresponds to the singular
zeroth-order solution.}
\label{f2}
\end{figure}

%

The contracting fixed points with the negative sign in eq. (\ref{48})
correspond to the time-reversed solutions describing a decelerated,
post-big bang contraction, which is also smoothly connected to the
perturbative vacuum. The general situation is illustrated in Fig. 3, which
shows the time behaviour of the Hubble factor $\bp$ (in units $k\ap=1$)
for the case of $d=3$ spatial dimensions. The dashed curves represent
the various branches of the singular zeroth-order solution \cite{5}:
\beq
\sqrt d \bp =\pm |t|^{-1}, ~~~~~~~~~~~~~ \fbp =\pm |t|^{-1},
\label{49}
\eeq
evolving towards ($(a)$ and $(c)$) or from ($(b)$ and $(d)$) the
singularity, expanding ($(a)$ and $(b)$) or contracting ($(c)$ and $(d)$).
The position of the singularity has been made to coincide with the origin
of the time axis, which thus separates the pre-big bang ($t<0$) from the
post-big bang ($t>0$) configurations. The four dashed
curves are related by T-duality and time-reversal 
transformation as follows:
\bea
{\rm T-duality} &:& ~~~~~~~~~ (a) \Longleftrightarrow (c), ~~~~~ (b)
\Longleftrightarrow (d) . \nonumber \\
{\rm t-reversal} &:& ~~~~~~~~~(a) \Longleftrightarrow (d), ~~~~~ (b)
\Longleftrightarrow (c) .
\label{410}
\eea

The numerical integration shows that, at least according to the model
(\ref{43}), only expanding pre-big bang and contracting
post-big bang configurations are regularized, to first order in $\ap$, as
illustrated by the solid curves. In the context of a theory that is
exactly duality-invariant, one may expect, however, a more symmetric
situation in which the symmetry pattern of the zeroth-order 
solutions is maintained  after regularization.

\begin{figure}
\centerline{\includegraphics[scale=0.8]{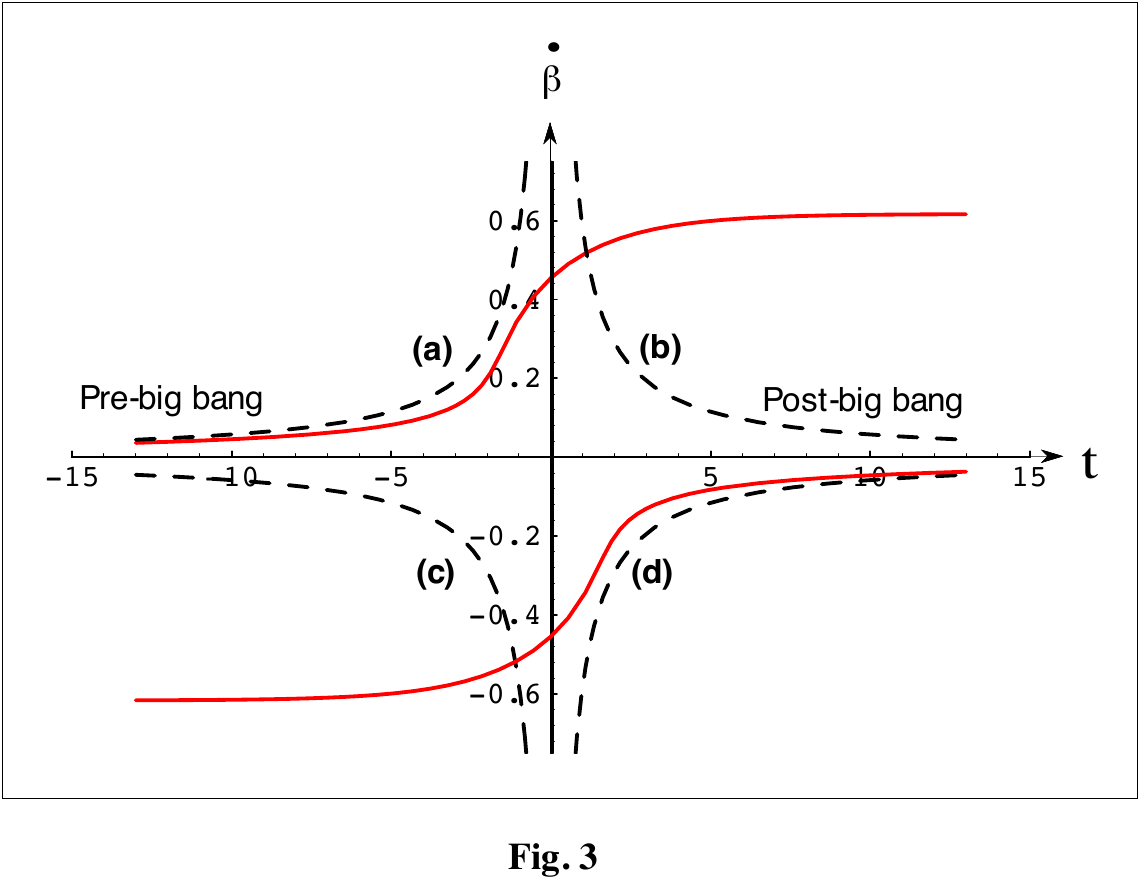}}
\caption{First-order $\ap$ corrections (solid curves) to the
branches of the singular zeroth-order solution (dashed curves).}
\label{f3}
\end{figure}


Since in the present example
the dual of the expanding pre-big bang branch is not regularized,
no smooth monotonic evolution from growing to decreasing curvature
is possible, unlike in models where one-loop corrections
are included \cite{10}--\cite{13}. In the loop case, however, the final
state of the background tends to remain in the pre-big bang sector
with $\fbp>0$, $\bp>0$, because of the final growing rate of the dilaton.
The expanding fixed point 
determined by the $\ap$ corrections corresponds instead
to a final configuration of the post-big bang type, with $\fbp<0$,
$\bp>0$, as illustrated in Fig. 4 by a numerical integration of the $d=3$
field equations. This means that, unlike what happens in the one-loop
model of \cite{13}, it is not impossible for the background to be
attracted by an appropriate potential in the expanding, frozen-dilaton
state of the standard scenario.

\begin{figure}
\centerline{\includegraphics[scale=0.8]{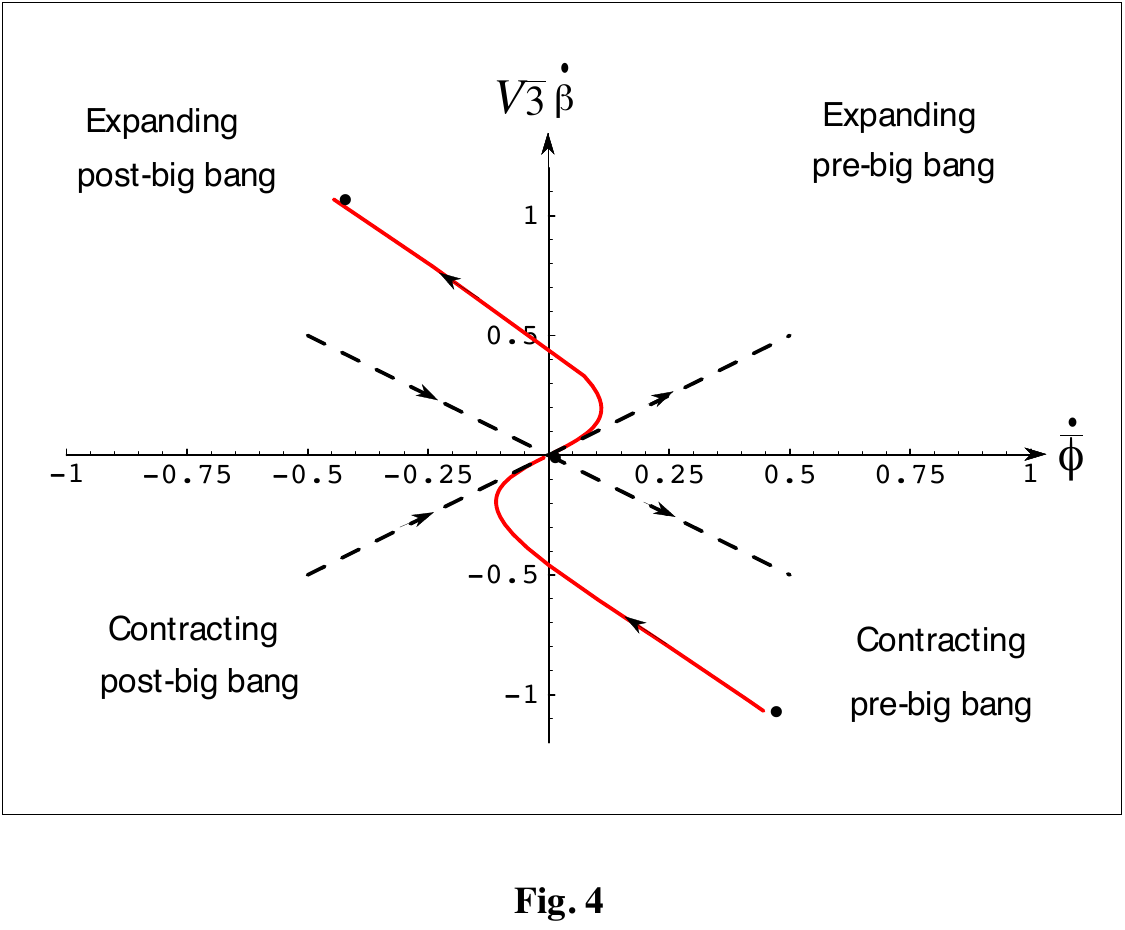}}
\caption{Evolution from the perturbative
vacuum to the expanding fixed point (and
its time-reversal) to first order in $\ap$. The dashed lines represent
the zeroth-order pre- and post-big bang solutions.}
\label{f4}
\end{figure}


Let us finally stress, to conclude the discussion of our example, that if
we start with sufficiently anisotropic initial conditions, the
background in general evolves towards a curvature singularity. 
However, the
isotropic fixed points
may also attract anisotropic backgrounds, and 
the set of anisotropic configurations
that are eventually attracted by the isotropic fixed point spans a region
of finite size in the space of initial conditions.  
This property of the first-order action (\ref{43}) is illustrated in Fig. 5,
where we have plotted various curves $\dot\ga (t)$ and
$\dot\ga (\fbp)$ obtained through a  numerical integration 
by varying the initial conditions of $\dot\ga$,
at fixed initial conditions for $\bp$. The plots refer to the case $d=2$,
$n=1$ and $\bp>0$, but they are qualitatively the same for any $d$ and
$n$, and can obviously be symmetrically extended to the plane $\bp <0$.
When the initial conditions are inside the area 
spanned by the curves shown in the figure, 
the curvature is
isotropized and regularized, otherwise the singularity is not avoided.

\begin{figure}
\centerline{\includegraphics[scale=0.8]{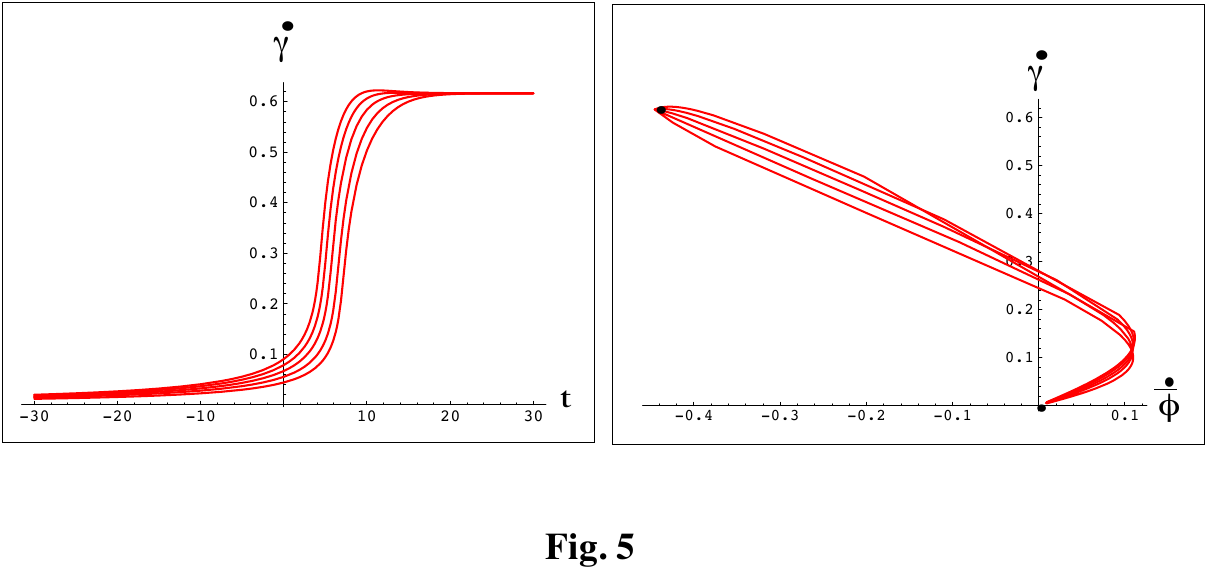}}
\caption{Attraction basin of the fixed points, in the space of
anisotropic pre-big bang initial conditions, for the case $d=2$, $n=1$, 
and for a particular fixed initial value of $\dot\beta$.}
\label{f5}
\end{figure}


The size of the attraction basin depends of course on the details of the
first-order action. By using, for instance, the $d=3$ effective action
obtained by transforming from the Einstein to the string frame the
Gauss--Bonnet invariant, we have found a much larger attraction basin
than the one illustrated in Fig. 5, so large as to include part of the
region $\dot\gamma <0$, namely contracting initial conditions. The lack of a
complete isotropization is not a negative aspect of the model since, in
a higher-dimensional background, the cosmological evolution should
indeed implement an effective dimensional reduction by separating
three expanding dimensions from the contracting ``internal" ones. 

\renewcommand{\theequation}{5.\arabic{equation}}
\setcounter{equation}{0}
\section {Conclusions}

The evolution of a cosmological background from the string 
perturbative vacuum may lead to a regime in which higher-derivative
contributions to the string effective action are large, while string 
loop effects are still negligible.  We have considered, in that regime, the
possible existence of a ``string-phase", with the background
curvatures frozen at a scale controlled by the string length parameter
$\la_s$, corresponding to an exponential evolution of the scale factor
and a linear evolution of the dilaton (in cosmic time). We have shown
that solutions of this kind  may represent an exact solution (to all
orders in $\ap$) of the tree-level action, and we have discussed an
explicit example (to first order in $\ap$) in which all expanding isotropic
backgrounds, evolving initially from the perturbative vacuum, are
necessarily attracted to that constant curvature state.

The main purpose of this paper was to point out the importance 
of $\ap$corrections for a singularity-free cosmology: by implementing a
mechanism of limiting curvature, they can regularize cosmological
backgrounds even in the absence of quantum loop effects. 
The emergence of
such a high-curvature string phase, in the weak coupling regime, leads
to a cosmological scenario rich of interesting phenomenological
consequences.

In a string-theory context, however, the properties of a background
obtained to first order in $\ap$ (and to any finite order in 
the $\ap$ expansion) are characterized by a certain degree of
ambiguity
\cite{6a}, because they are not  invariant under field
redefinitions. A truly unambiguous cosmological background should
correspond to the solution of an exact conformal field theory
\cite{28}, which automatically includes all orders in $\ap$. 
In addition, the quantum back-reaction of loops and radiation, as well
as an appropriate non-perturbative dilaton potential, are certainly
required to complete the transition from the string phase to the
radiation-dominated, constant dilaton phase of the standard scenario.
In this sense, the results of this paper are still preliminary to the
formulation of a complete and realistic string cosmology scenario.
Nevertheless, they confirm previous conjectures, clarify the relative
role of $\ap$ and loop corrections, and motivate a more systematic
study of higher-order and exact conformal solutions with constant
curvature and linear dilatonic evolution.

\vskip 2 cm

\section*{Acknowledgements}

We are grateful to Krzysztof Meissner and
Arkady Tseytlin for  helpful discussions. M. G. and G. V. are
supported in part by the EC contract No. ERBCHRX-CT94-0488.


\end{document}